# Global Distribution of Water Vapor and Cloud Cover—Sites for High Performance THz Applications

J. Y. Suen, *Student Member, IEEE*, M. T. Fang, and P. M. Lubin

*Abstract*—Absorption of terahertz radiation by atmospheric water vapor is a serious impediment for radio astronomy and for long-distance communications. Transmission in the THz regime is dependent almost exclusively on atmospheric precipitable water vapor (PWV). Though much of the Earth has PWV that is too high for good transmission above 200 GHz, there are a number of dry sites with very low attenuation. We performed a global analysis of PWV with high-resolution measurements from the Moderate Resolution Imaging Spectrometer (MODIS) on two NASA Earth Observing System (EOS) satellites over the year of 2011. We determined PWV and cloud cover distributions and then developed a model to find transmission and atmospheric radiance as well as necessary integration times in the various windows. We produced global maps over the common THz windows for astronomical and satellite communications scenarios. Notably, we show that up through 1 THz, systems could be built in excellent sites of Chile, Greenland and the Tibetan Plateau, while Antarctic performance is good to 1.6 THz. For a ground-to-space communication link up through 847 GHz, we found several sites in the Continental United States where mean atmospheric attenuation is less than 40 dB; not an insurmountable challenge for a link.

## I. Introduction

FROM the early days of terahertz science and technology, it has been known that strong atmospheric absorption, almost all from water vapor, is perhaps the most significant challenge in exploiting this spectrum. For many applications, this challenge is insurmountable; there are no practical ways of changing atmospheric water vapor, nor can the fundamental physics of molecular interaction be avoided.

In this paper we concentrate on studying the worldwide distribution of water vapor and its effects on both THz radio astronomy and communications. In radio astronomy, one is often looking for specific spectral features in distant sources, such as high redshift galaxies, or the remnant radiation from the early universe and many times this means there is a loss of information if one were to switch to a lower frequency. One solution is to go to a space telescope, such as the recent Herschel and Planck observatories. However, space systems are very costly with long development times and an inherently limited lifetime, with compromises made in limited aperture sizes and instrument suites. Ground based telescopes such as the Atacama Large Millimeter/Submillmeter Array (ALMA), currently under construction, and the proposed Cerro Chajnantor Atacama Telescope (CCAT) are situated in very high and dry plateaus. For example, ALMA consists of an array of fifty-four 12-meter and twelve 7-meter reflector antennas, and CCAT is proposed to be 25 meters in diameter. Both of these are located in the Atacama Desert in Chile at 5 Km and 5.6 Km altitude, respectively. Other similar projects include the South Pole Telescope, and the authors currently operated microwave through THz instruments at the Barcroft Observatory on White Mountain, California, USA at 4 Km altitude. Observing sites are chosen for a variety of reasons that often require compromises. Good atmospheric conditions are always desired but so are good access and basic, economical, infrastructure. These inevitably lead to real tradeoffs in performance.

These sites were individually located and often painfully characterized, focusing on astronomical observatories and often over short periods of time [1-6]. Satellite data has been used in these studies, over small regions and with limited resolution and timespans [7-9]. In this paper we look at the whole Earth over the period of one year with kilometer-level resolution. In this way we can explore vast areas rapidly and at low cost without mounting expeditions. We also study sites that are known to be excellent as well as largely unknown sites on an unbiased and completely global basis. High resolution is necessary in order to identify geographically small areas, such as mountain peaks, which may be locally dry and year-round coverage necessary to be properly representative of long-term performance.

To accomplish this, we analyzed the complete water vapor data from two NASA Earth Observing System (EOS) satellites over the 2011 calendar year. Combining this data with our MODTRAN 5 models, we found global atmospheric attenuation in the atmospheric THz windows from 275 to 1,500 GHz, the band at which site selection has the most

This research was supported by NASA California Space Grant, grant number NNX10AT93H. We acknowledge support from the Center for Scientific Computing at the CNSI and MRL: an NSF MRSEC (DMR-1121053) and NSF CNS-0960316.
J. Y. Suen is with the Department of Electrical and Computer Engineering, University of California, Santa Barbara, CA 93106 USA (e-mail: jsuen@ece.ucsb.edu).
M. T. Fang and P.M. Lubin are with the Department of Physics, University of California, Santa Barbara, CA 93106 USA.

significant impact. With this data and methodology in hand, as well as the machinery to analyze this data we can undertake similar analysis from radio into the UV in future papers, but for the purpose of this paper we focus on the THz windows.

We present our data in two ways. If we wish to consider a THz satellite communications link, one that exploits the very high potential data rate available in the THz regime due to the wide bandwidths available, we present the mean (as distinct from a median) atmospheric attenuation for a 45 degree slant path. This represents a typical link from a mid-latitude location to a geostationary satellite and can be used to find the long-term average data link rate using a link budget calculation. For the second case of a site for astrophysical observations in the THz regime we also show the median excess integration time factor, that is, the additional integration time necessary to reach a given SNR, caused solely by atmospheric water vapor and clouds. This would be of interest in siting a sub-mm or THz telescope as it relates to the overall observational productivity of such an instrument, and provides not only a natural comparison between sites, but can also be compared with a spaceborne instrument. In a recent article [10] we compared the various issues, particularly noise sources and telescope tradeoffs in the far IR (0.1 to 10 THz), which is also useful in this context for astronomical applications.  Our analysis allows us to study the distributions (e.g. various PWV percentiles) and temporal distribution of each point on the surface of the Earth. We can observe the PWV "weather" patterns over the course of the year as well. We further modeled atmospheric radiance due to water vapor which we present as an antenna temperature. We present color maps depicting combinations of scenarios, slant path, and the 8 atmospheric windows in this article. High-resolution versions are available in the online appendix. In this paper, we first discuss the data, our models, and then draw conclusions on possible sites for THz astronomy and satellite communication links.

## II. DATA PROCESSING AND MODELING

Atmospheric water vapor is expressed as precipitable water vapor (PWV), sometimes also referred to as total precipitable water (TPW).  PWV represents the total water content in a column of the atmosphere, from ground level to the top of atmosphere, if it was completely converted into liquid water. Water in the form of ice is not included in this metric. This is quite reasonable as the absorption bands for ice are very different and generally far less absorptive than the liquid phase. The units of PWV are length, i.e. $mm^3/mm^2 = mm$ which is the height of the equivalent liquid level if the water vapor were condensed on the ground.

We utilized the complete 2011 calendar year dataset from the Moderate Resolution Imaging Spectrometer (MODIS) [11].. MODIS senses 36 channels from visible through long-wave IR with at worst 1-km resolution, and independent units are flown on two NASA EOS satellites, Aqua and Terra. The high resolution of the instrument coupled with global land and sea coverage makes MODIS data unique. Sensed radiances were processed by the MODIS Atmosphere team with a synthetic regression algorithm into the Atmospheric Profile product, known as MOD07_L2 or MOD07[1], which includes total column PWV, determined primarily from three bands centered at 6.7, 7.3, and 8.6 μm [12]. A 5 by 5 square of infrared radiances are combined to a single 5 km PWV pixel, with coverage over the entire surface of the Earth. Both satellites are in a sun-synchronous orbit, providing coverage both in the morning and afternoon local time, and make nighttime passes 12 hours later. Compared to other microwave or infrared humidity sensing systems, MODIS had the highest spatial resolution and also provided four passes per day, reducing bias from the typically afternoon passes of meteorological satellites.

The PWV cannot be retrieved through clouds, and is provided only when 9 of the 25 MODIS samples are cloud free and a flag indicates when this condition was not met. The MOD07 algorithm relies on the MOD35 cloud mask algorithm, which uses 20 infrared and visible MODIS bands along with terrain and sun data to provide an indication of cloud cover. Clouds are sensed by primarily their high reflectance and low temperature [13]. We used the cloud data to generate cloud cover statistics. Detection of clouds by the MOD35 algorithm is dependent on optical thickness. Thin clouds, specifically thin cirrus will not be flagged. Due to the additional water vapor and scattering associated with clouds [14], we assumed that THz transmission would not be possible through a pixel declared cloudy. Thus our analysis is centered around the clear-sky PWV, though we also indicate the number of cloudy days.

TABLE I
MOD07 PWV COMPARED WITH INDIVIDUAL SITE CHARACTERIZATIONS

| Location | MOD07 Median PWV (mm) Clear-Sky | MOD07 Median PWV (mm) Estimated All-Weather | MOD07 Cloudy Sample Fraction | Site Measured All-weather Median PWV (mm) |
|---|---|---|---|---|
| Dome A, Antarctica | 0.07 | 0.08 | 0.138 | 0.14[1], 0.22[1] [15, 7] |
| Dome C, Antarctica | 0.10 | 0.11 | 0.112 | 0.24[1], 0.28[2], 0.75[2,3] [15, 16, 4] |
| South Pole | 0.14 | 0.20 | 0.267 | 0.32[1], 0.42 [15, 17] |
| ALMA Site, Chajnantor | 1.37 | 1.86 | 0.252 | 1.1[3,4] [2] |
| Barcroft Observatory[7] | 4.22 | 7.60 | 0.310 | 1.75 [1] |
| CARMA Site | 5.56 | 7.16 | 0.198 | 5[5] [6] |
| Mauna Kea[7] | 14.55 | 21.84 | 0.284 | 1.5, 1.5[6] [15, 18] |

MOD07 data is over the entire year of 2011. Site medians are annual unless noted. All-weather estimates assume cloudy samples have infinite PWV. Refer to Table IV for site altitudes and coordinates.
1. Austral Winter only
2. Observations over ~2 months during the Austral Summer
3. Observations made only while telescope operating (implies clear-sky or better conditions only)
4. Observations over 5.5 years (May 2006-Dec.2012)
5. Observations over approx. 13 of 16 months (Oct. 2002-Feb. 2004)
6. Daytime only
7. Site is surrounded by steep terrain with large PWV variations

---

[1] Collection 5.1 data was used. (Refers to a specific version of MODIS retrieval algorithms)

*A. Data Validation*

The authors of the MOD07 algorithm compared retrieved PWV data to ground measurements ground measurements at a mid-latitude site. Over approximately 6 years of data, it was found that MOD07 had an average wet bias of 0.7 mm for the morning pass and 0.3 mm for the afternoon pass, when considering only dry measurements (<15 mm PWV). The bias was less than 0.1 mm wet when compared to GOES and 0.6 mm dry when compared to radiosonde measurements. [12]

Comparing MOD07 PWV data to several site-specific measurements (Table I, Fig. 1) indicated a reasonable match. It is clear that relative PWV trends match well both in Antarctica and through the mid-latitude sites. We note that many site-specific measurements, based on millimeter-wave radiometers or radiosonde measurements do not distinguish between clear-sky and inclement weather. For example, all of the sites listed are subject not only to cloud cover, but snowstorms. These conditions may not result in a dramatic PWV increase, but can preclude observations, especially for large arrays, which rely on atmospheric phase stability.

All-weather PWV values were estimated by assuming cloudy PWV samples had infinite PWV. This should result in an overestimation of all-weather PWV since clouds have finite PWV. While absolute errors lead to larger deviations at low PWV, mainly concerning the extreme levels in Antarctica, the difference is acceptable in the other dry areas of interest.

For example, measurements of the ALMA and Combined Array for Research in Millimeter-wave Astronomy (CARMA) site, which lie in the range from 1-5 mm, are close to the measured ground data. It is important to keep in mind that measured ground data often refers to very different time periods that are subject to long term atmospheric trends (El Nino etc.) and that neither site-specific nor MOD07 data presented was contemporaneous. Figure 2 compares cumulative distributions of PWV between the MOD07 data and the site measurements of [2]. Distributions are close, with the MOD07 data being slightly wetter. We note that the site measurements were taken over 5.5 years (May 2006-Dec 2012) versus the single (2011) year of MOD07 data analyzed. Additionally, the data in [2] was only taken during operation of the APEX telescope and therefore should be comparable or even drier than MOD07 clear-sky data. Figure 3 depicts the same MOD07 data in histogram and time series form, also showing cloudy samples.

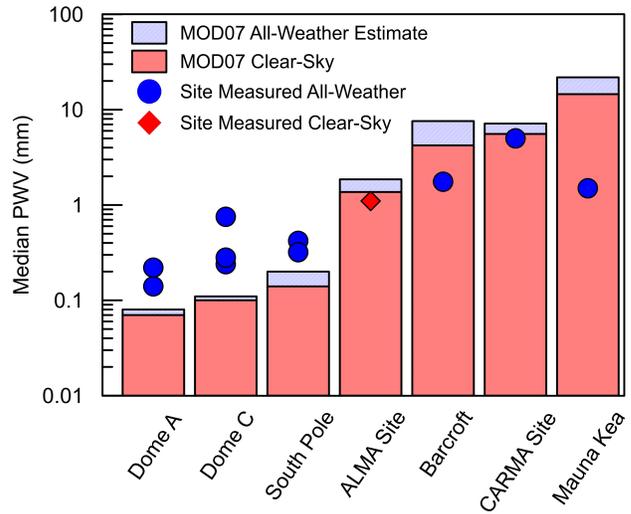

Fig. 1. Comparison of median PWV from MOD07 and site measurements. Except for sites surrounded with steep terrain with large increases in PWV (Barcroft and Mauna Kea), MOD07 data exhibits identical trends to site measurements. MOD07 data in the Antarctic is dry whereas over the mid-latitudes, MOD07 data is slightly wet.

For the case of Antarctica, we do note that measurement of such low PWVs is a significant challenge and can result in wide variations of even ground-based measurements, e.g. a reported median PWV at Dome C three times greater in reference [4] versus [15] due to different observation periods and methods. Nonetheless, MOD07 PWV measurements match the same qualitative trend as site-specific measurements. A further complication is that many site-specific measurements are conducted over limited periods of time and extrapolated to year-round performance. Satellite data results from direct sensing over the whole year and thus

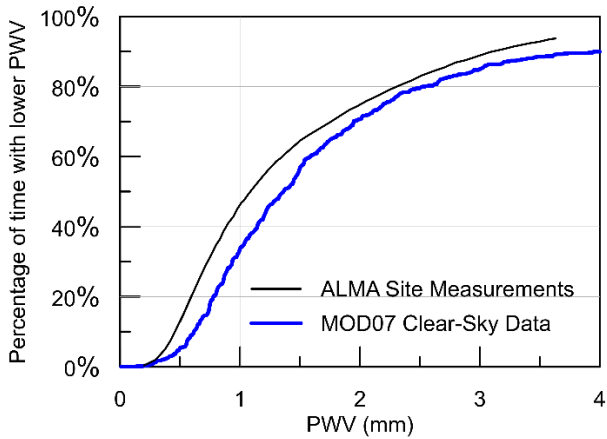

Fig. 2. Cumulative distribution comparing ALMA site measurements from Ref. [2] to MOD07 data (n=688). ALMA data is over 5.5 years while MOD07 data was over the 2011 calendar year. ALMA data was only collected while the APEX telescope was operating, implying, at minimum, clear-sky conditions. Note the excellent agreement between the satellite data and the ground characterization data.

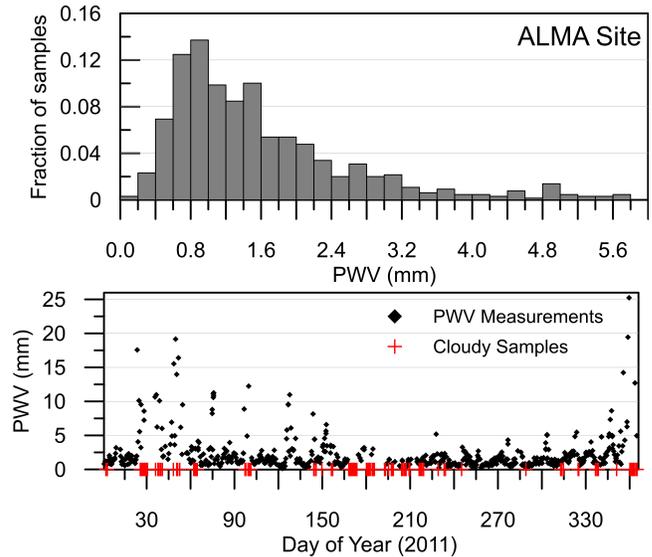

Fig. 3. MOD07 data of the ALMA site. The clear-sky PWV data is shown as a histogram (top). The time series (bottom) shows seasonal variations. 688 clear and 232 cloudy points are shown.

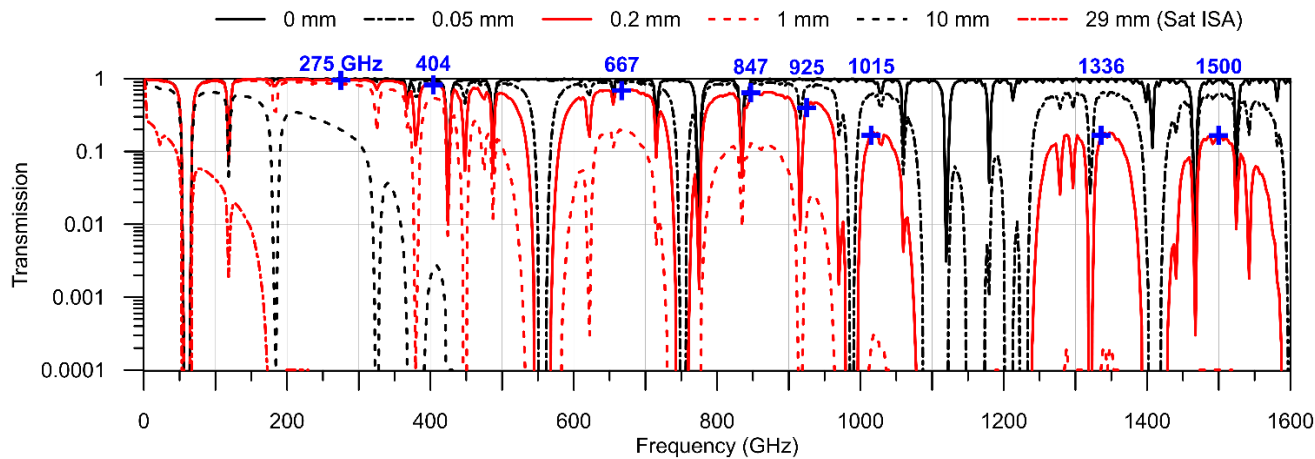

Fig. 4. Transmission versus frequency as modeled by MODTRAN 5 for a vertical path from space to sea level. Various levels of precipitable water vapor are depicted, from no water to a completely saturated International Standard Atmosphere. Frequencies used in modeling are also marked. Absorption bands of oxygen and other atmospheric gasses are seen in the 0 PWV case.

less biased.

As inherent in any optical instrument, the MODIS spectrometer has a beam that is not ideal or infinitesimally small and relies on propagated radiation through the atmosphere. We see that sharp mountain peaks show influence from lower, wetter surrounding areas as evident by the Barcroft Observatory and in particular the Mauna Kea measurements, both sites which are surrounded by steep terrain with large PWV gradients. As satellite water vapor retrieval techniques advance, we expect that analysis on such data will locate additional, smaller, sites.

*B. Data Processing*

The MOD07 data, as delivered, was not gridded; the pixel centers are not the same across different satellite passes. We gridded the data into pixels 0.05 degrees latitude and longitude (5.56 km pixels at the equator). Our data analysis found the mean clear-sky PWV, a number of percentiles for each pixel as well as the cloudy sample fraction, as defined by the number of points marked as cloudy divided by the total count of points, over the entire surface of the Earth. The time series data for each pixel could also be retrieved. Each gridded pixel contained a median value of 912 samples and the data appeared smooth.

The MOD07 data came formatted as 2.7 megapixel images each corresponding to 5 minutes of data acquisition. Over a full year and two satellites, there are a total of 288 billion data points over 210,000 files. Initial efforts attempted to sort the data points by location and time using geospatial indices in a relational database, but the inability to parallelize the processing and the memory necessary made this approach infeasible.

Instead, a program was written in MATLAB using the MapReduce distributed parallel processing model, which separates processing into a parallel mapping and a centralized reduction stage. In the mapping stage of the processing, a central process distributed data files to independent worker processes. Each worker process loaded and parsed an arbitrary subset of the files and iterated through each image pixel, applying data validation logic. In order to compute the mean, these mapping processes kept, keyed at the geographic coordinates, an individual running sum of PWV data, valid, and cloudy pixel counts as values. When the mapping stage was concluded, the reduction stage received the data from each mapping process and then combined the data by simply summing the values over each final grid location. The mean clear-sky PWV and the cloudy sample fraction was then calculated at each grid location. For percentile computation, the parallel mapping processes performed file parsing and data validation but passed the time series data to the reduction stage which computed the percentiles at each grid location independently. With the much smaller 26 megapixel output maps, further processing was then performed on a desktop computer, namely applying the integration time model and generating output graphics.

We utilized the Knot cluster at the UC Santa Barbara Center for Scientific Computing. A full run of the data consumed in total 144 CPU core-hours, running on 2.0 GHz Intel Xeon X7550 processors, and 372 GB of RAM. The 716 GB of file data was stored on a HP IBRIX X9320 system, connected via Infiniband. With 12 worker processes, a real-time speedup of 10.9 was observed, suggesting that processing was just beginning to be limited by file access rates (most parallel filesystems incur significant per-file access overhead).

*C. Phenomenology and Modeling*

In order to quantify the effects of water vapor on THz systems, we first created an accurate and computationally efficient model of PWV on atmospheric transmission and radiance. Due to its highly polar nature, water is the primary absorber in THz atmospheric propagation. The large number of inter- and intra-molecular vibrational modes results in large absorption bands. In contrast, non-polar and weakly polar molecules such as $O_2$ and $CO_2$ result in narrow absorption lines, which can usually be avoided.

Water is heavily concentrated at lower altitudes because of its high freezing point relative to other atmospheric gasses. The partial pressure of water vapor, and hence its concentration, shows a high temperature dependence; approximately exponential with temperature. Additionally, water vapor enables a positive feedback loop, since it reflects mid-IR radiation. Thus, high-altitude and naturally cold sites exhibit very low amounts of water vapor.

In addition to absorption, atmospheric gasses emit radiation due to heating from the Sun and surface of the Earth. This radiation can be scattered, multiple times, by aerosols in the atmosphere. As a result of the necessarily warm temperatures of water vapor and the wide absorption bands caused by the large number of inter- and intra-molecular modes, water vapor forms the primary component of atmospheric radiance, typically 1-2 orders of magnitude greater than that of remaining atmospheric gasses.

We considered 275, 404, 667, 847, 925, 1015, 1336, and 1500 GHz frequencies, which were selected to be approximately in the middle of each atmospheric window while being away from absorption line features (Fig. 4). These frequencies span the THz region of interest: the impact of water vapor is relatively lower below 275 GHz. Above 1600 GHz, severe attenuation rules out use anywhere but the driest parts of Antarctica. Attenuation coefficients as a function of PWV were found using the MODTRAN 5.2 Mid-latitude Winter model (PcModWin5 v1.3.3, Ontar Corp., North Andover, MA, USA) and we found attenuation was very well modeled ($R^2 > 0.9995$) with the Beer-Lambert law. We used this simple logarithmic model to avoid running a full MODTRAN model for every single point.

We also modeled atmospheric radiance in the multiple scattering mode of MODTRAN. Nearly all of the atmospheric irradiance at THz frequencies is due to thermal emission by the atmosphere, where atmospheric gasses radiate energy after being heated by absorption of radiation from the Sun and the Earth. Where PWV, and hence THz absorption is high, the atmosphere will have an emissivity close to 1 and the antenna temperature saturates at the physical temperature of the low-level air, 294 K in the model used. Radiance also is caused by both direct and scattered radiation emitted by the Sun. We assumed the Sun was at 45 degrees in elevation, and aligned in azimuth. At the long wavelengths being studied, the directionality, dependence on solar angle, or even the difference between day and night is minimal since the majority of solar radiation is visible, with very low amounts longer than 5 μm. We fit the radiance accurately using a piecewise-cubic interpolation function and used this as a computationally faster model. (The Sun is usually the driving factor for the day/night variations of PWV, however, with nighttime often far superior to daytime and with Sunrise and Sunset often having the most temporal variation.)

We assumed a transmission path from the top of the atmosphere to sea level to avoid the need for ground elevation data. Temperatures and pressures were also assumed to be a mid-latitude winter atmosphere. Fig. 5 compares the assumed model for the South Pole to the MODTRAN model with only the site-specific altitude, as well as a model with a radiosonde measured atmospheric profile (scaled to match PWVs). As can be seen, the assumed model matches the radiosonde profile well to 1 THz, and shows small differences through 1.6 THz. In contrast, application of only the altitude to the winter atmosphere profile results in higher transmission than either of the two models.

We also considered models other than MODTRAN that are in common use by the THz astronomy community. We compared both the *am* mid-latitude generic atmosphere [18] and ATM [19] models with MODTRAN. Figure 6 compares *am* and MODTRAN at sea level. Where there is no atmospheric water vapor, *am* shows approximately 3 dB more attenuation at 1.2 THz than MODTRAN due to the dry air collision-induced absorption which is not modeled in MODTRAN. However by 500 μm PWV, the two models produce very similar results except in the deep absorption lines and troughs. ATM showed a similar trend but with some variance in attenuation. Compared with *am,* ATM had roughly 2 dB less attenuation over most of the band for both PWV cases. For the PWV values and frequencies we are primarily interested in, we believe there is no practical difference between these models. *Am* and similar models are able to operate with a much finer resolution than MODTRAN's 3 GHz apodized resolution, which is important for modeling very high resolution spectrographic astronomical observations. MODTRAN is able to model up through the ultraviolet

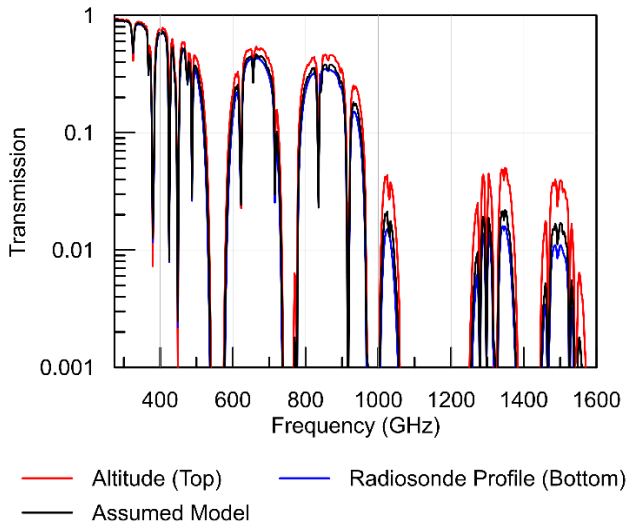

Fig. 5. Comparison of South Pole MODTRAN transmission spectra for various assumptions made. All spectra are at 0.42 mm PWV. Our assumed model uses sea-level altitude and atmospheric profile, only using site-specific PWV. The effects of using the site specific altitude (2,835 m) and a radiosonde atmospheric profile is shown. Differences between the assumed model and the radiosonde profile are minimal under 1 THz.

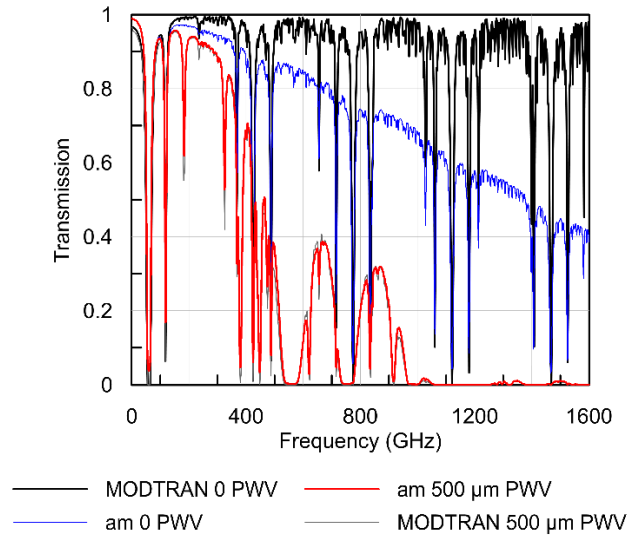

Fig 6. Comparison between the *am* and MODTRAN models. Due to the dry air collision-induced absorption only modeled in *am*, attenuation differs by ~3 dB in the 0 PWV case. However, by 500 μm PWV, differences are minimal for the bands and spectral resolution under consideration.

spectrum while *am* and ATM are used below 2 THz in general.

*D. Integration Time Factors*

While determining atmospheric loss and radiance is useful in itself, we sought a metric which could directly relate the effects of atmospheric loss and cloud cover on the ultimate usability of a site. This is best measured in terms of integration times, which expresses both the productivity of an astronomical observatory and the data rate for a communication link.

For modeling a long-term observation, we utilized the integration time relationship, where the integration time $\tau$ for a given SNR is

$$\tau = \left(\frac{SNR\, P_n}{\sqrt{2}\, P_s}\right)^2, \tag{1}$$

where $P_s$ is the signal power units of watts and $P_n$ the noise power in units of $W/\sqrt{Hz}$ (where Hz is bandwidth, i.e. $P_s$ is noise-equivalent power). We take equation 1 as our baseline case and apply two factors. The mean atmospheric transmission $T$ simply decreases the signal power while keeping noise power constant. This assumption neglects the effect of atmospheric radiance, which will be discussed later. Also, we do not include extraterrestrial backgrounds in this paper.

We further assume that no observations can be made when it is cloudy, and therefore integration time is increased by $(1-C)^{-1}$, where $C$ is the cloudy sample fraction. As previously discussed, the cloud mask algorithm relies on the optical thickness of clouds and will not flag optically thin clouds. We therefore assumed that flagged clouds are optically thick in the visible and infrared and will also preclude THz transmission.

If we define the excess integration time factor, $E_\tau$, as the ratio of the integration time with transmission and cloud effects $\tau_n$ to the integration time without, $\tau$, we have,

$$E_\tau = \frac{\tau_n}{\tau} = \frac{1}{(1-C)T^2}. \tag{2}$$

Under this definition, we assume that noise sources, such as receiver noise and thermal mirror emissions are identical. As discussed later, this does not directly compare a ground system with a spaceborne system because in practice the design of both systems would not be identical. Instead it provides a good comparison baseline, discussed later. See our recent paper in reference [10] for a much more complete discussion of the various issues relating to observing weak astronomical sources through the atmosphere and with extraterrestrial backgrounds included.

Since the effects of PWV are exponential, the ~2 orders of magnitude difference in PWV over the surface of the Earth will translate to very dramatic changes in THz absorption; a factor of two in PWV may be a factor of 100 or more in transmission in the THz bands. Additionally, integration time scales to the square of this exponential absorption factor, whereas it scales proportionally to increasing cloud cover. Therefore, in most cases the importance of PWV will be far greater than that of cloud cover.

To this point, we have not considered the effects of atmospheric radiance. A difficulty in analysis is that due to the inherent dependence of this noise power on observation bandwidth, aperture size and acceptance angle, which are all specific to the system under consideration. Further, while atmospheric transmission will dominate the signal attenuation of nearly any ground-based THz system, the relative effects of radiance will depend upon other system noise sources, particularly thermal antenna radiance and receiver noise. As a result we present radiance as an antenna temperature. In the terahertz regime the Raleigh-Jeans limit applies and thus for near-diffraction limited systems the equivalent noise power for an antenna temperature,

$$P_T = k_b \Delta \nu\, m\, T_a, \tag{3}$$

for Boltzmann's constant $k_b$, bandwidth $\Delta \nu$, $m$ accepted modes (1 or 2 polarizations for a diffraction-limited system), and antenna temperature $T_a$.

Current uncooled, coherent, solid-state THz receivers that are likely to be used in communication applications have noise temperatures in the hundreds to thousands of kelvin, while cryogenic bolometers and SIS receivers, a common choice for astronomy, can be in the range of tens of Kelvin or less. We also must consider the specific source. For example, a solar observation is inherently limited by the 5,800 K temperature of the source and therefore the BLIP limit (shot noise) of the source will dominate over the noise of the atmosphere and any reasonably designed receiver.

In short, unlike attenuation, the effects of atmospheric radiance are completely dependent on the source and system design and can dominate, be negligible or some level in between. Again, our recent paper in [10] discusses this in detail. We present the antenna temperature associated with radiance that may be used with equations 1 and 3 to calculate the effects.

### III. RESULTS AND ANALYSIS

The large number of frequency bands, possible scenarios and data presentation precludes a full reproduction of results here. High-resolution versions of the maps shown here are available on the online supplemental appendix and the reader is encouraged to contact the authors for further study of a specific scenario. In the analysis that follows, we focus on the best sites for all applications, the productivity of astronomical telescopes at these sites, and an analysis of suitability for space communication applications. To this extent, we first present raw PWV and cloud data and show the direct impact on median transmittance and the atmospheric radiance antenna temperature for several frequency windows. We then apply our integration time factor to the PWV data in an astronomical scenario (zenith path). Finally, we utilize mean PWV data to analyze a THz ground-to-satellite communication link by showing transmission at a 45° path.

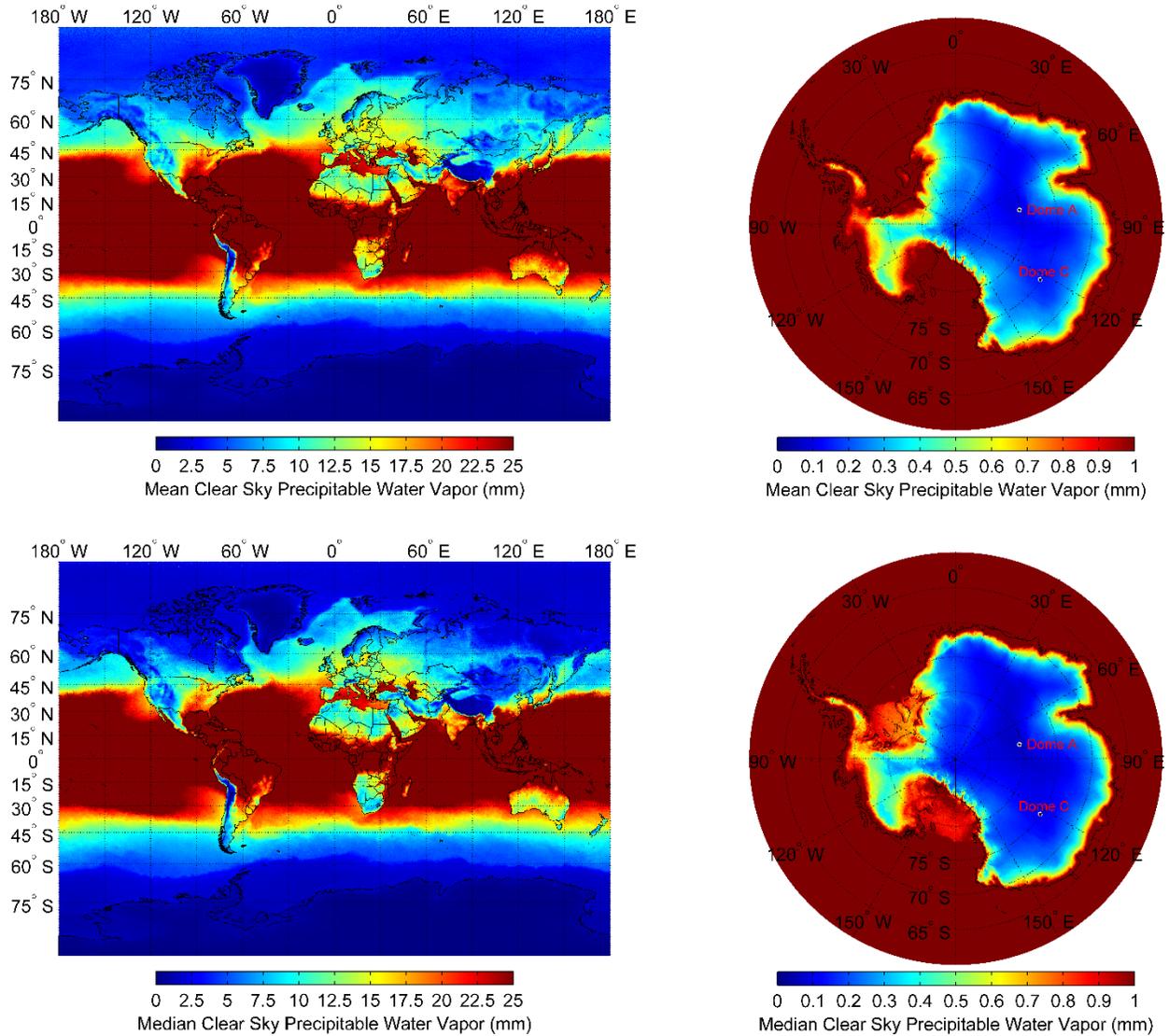

Fig. 7. Global distribution of mean and median clear-sky PWV, derived from combined MOD07 Aqua and Terra satellite infrared data for the 2011 calendar year. Higher resolution color versions of all maps in this paper are included in the supplemental figures, available online.

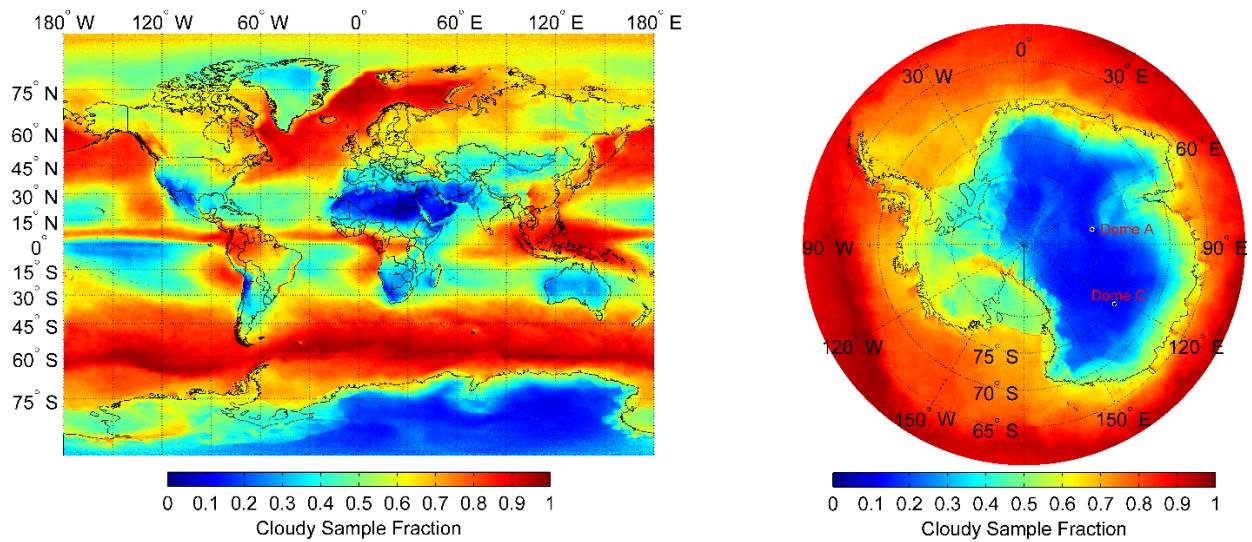

Fig. 8. Global distribution of cloudy sample fraction, derived from combined MOD07 Aqua and Terra satellite infrared data, for the 2011 calendar year.

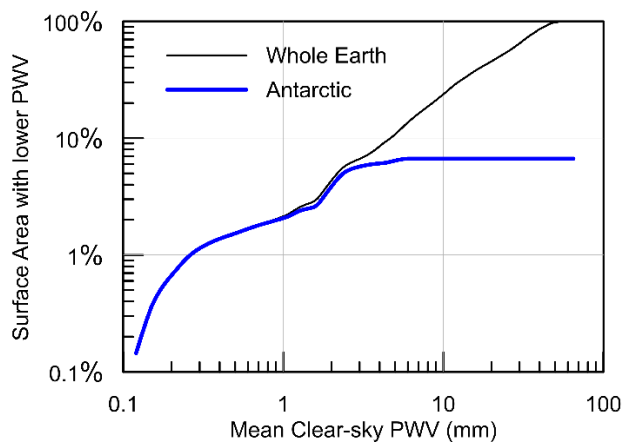

Fig. 9. Distribution of mean clear-sky PWV by surface area. Contribution from Antarctic regions (latitude <60° S) is shown. Total surface area is 510 million km$^2$.

Figures 7 and 8 depict global median (50$^{th}$ percentile) and mean clear-sky PWV and cloud fraction, as raw input data to our models. Immediately evident should be the low PWV areas of the entire Antarctic continent, the Andes, the Greenland Ice Sheet, and the large area of the Tibetan plateau. Areas which have significantly low PWVs are the Alaskan mountain range, the ranges of the Western United States, the Scandinavian mountains, and the Altai Mountains between Russia, Mongolia and China.

Figure 9 depicts the cumulative distribution of the median clear-sky PWV as a function of surface area, with the Antarctic region (>60° S latitude) separated. Fundamentally, all area under 1 mm PWV is Antarctic, while 99% of the remaining area sits above 3.0 mm. We also see that a significant amount of non-Antarctic surface area, 20% is drier than 10.6 mm and 4.7% is drier than 5.0 mm. From the data, expect that optimal Antarctic sites will have a median clear-sky PWV under 1 mm while the best non-Antarctic sites will range from around 1-10 mm PWV.

In the case of extremely low level sites even within the Antarctic plateau, there is still a considerable difference between sites such as the South Pole, Dome A, Dome C, Dome F and Ridge A. Figure 10 compares time series between Domes A and C and the South Pole, showing the driest conditions at Dome A, followed by Dome C and then the Pole. The Antarctic seasonal trend is visible, with higher water vapor levels during the warmer austral summer.

*A. Astronomical Scenarios*

We first interpret the PWV data by studying models for a zenith transmission path. This is a best-case path and representative of astronomical observations, some of which can be scheduled when the target is approximately overhead. It is also is useful for communication links to non-geostationary orbits where the path is close to overhead.

Figs. 11-13 depict the median clear-sky transmission at the 404, 667, and 925 GHz windows for a zenith-pointing path. It can be observed from Fig. 11 that low attenuation, less than 20 dB, exists at many locations over the surface of the Earth. Fig. 12, at 667 GHz, shows a significant increase in attenuation. Transmission to sites in the mid-latitudes are significantly reduced while the Greenland Ice sheet, high Atacama Desert,

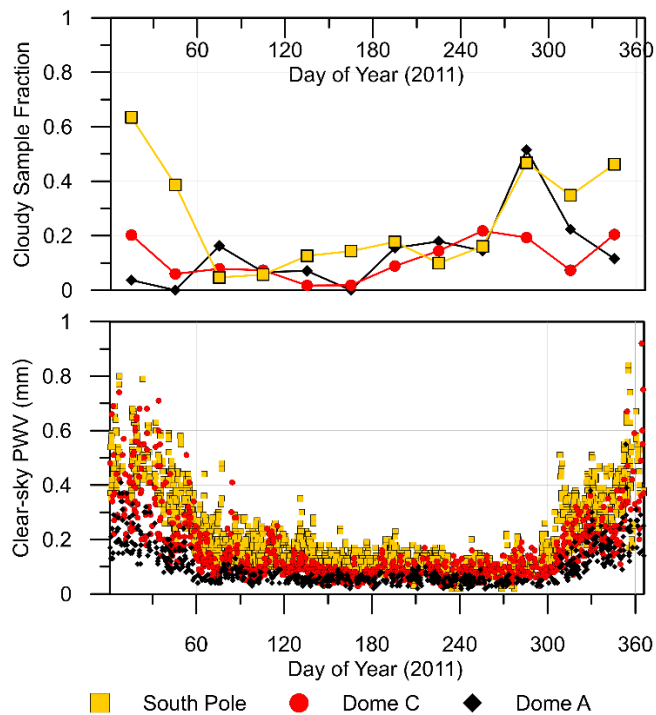

Fig. 10. Time series and cloudy sample fraction of three Antarctic sites: Domes A and C, and the South Pole. Note effects of the warmer austral summer (days 265 through 79) can be seen.

Tibetan Plateau and Antarctica remain excellent. By the 925 GHz window (Fig 13), only portions aforementioned sites remain above the -20 dB mark, with the remainder of the Earth worse than -100 dB. The benefits of the topography of Antarctica only become evident at the 925 GHz window and above. At the upper end of the frequency range, the 1336 and 1500 GHz windows are only accessible with 40 dB attenuation at prime sites, or with 3 to 10 dB attenuation in the interior of Antarctica.

Figures 14 and 15 show the effects of atmospheric radiance for 404 and 667 GHz vertical paths, respectively. As PWV increases, antenna temperature rapidly approaches the 294 K assumed ground level temperature of the MODTRAN model. At 404 GHz, this occurs at the moderate PWV of 16 mm, whereas this occurs at 6.3 mm at 667 GHz. Only the lowest PWV areas, typically Antarctic, have radiance which is significantly lower. The assumed ground temperature represents a standard atmosphere at sea level, hence is generally too high for high-altitude sites. At higher PWVs, the antenna temperature of atmospheric radiance is close to the ground air temperature, while for lower PWVs, mainly in Antarctica, this is often not the case and detailed atmospheric models are necessary.

Figure 16 depicts the 667 GHz integration time factor for a zenith path and with cloud cover incorporated; radiance is not considered. This scenario is closest to observations where the photon statistics of the source is the dominant (i.e. BLIP limited system). While an integration time factor of 1 represents no atmospheric effects, it does not compare directly to a space telescope. Ground-based instruments can easily have two orders of magnitude more collecting area (1 m versus 10+ m diameter), compensating for a $100^2=10^4$ larger integration time. With larger ground based system,

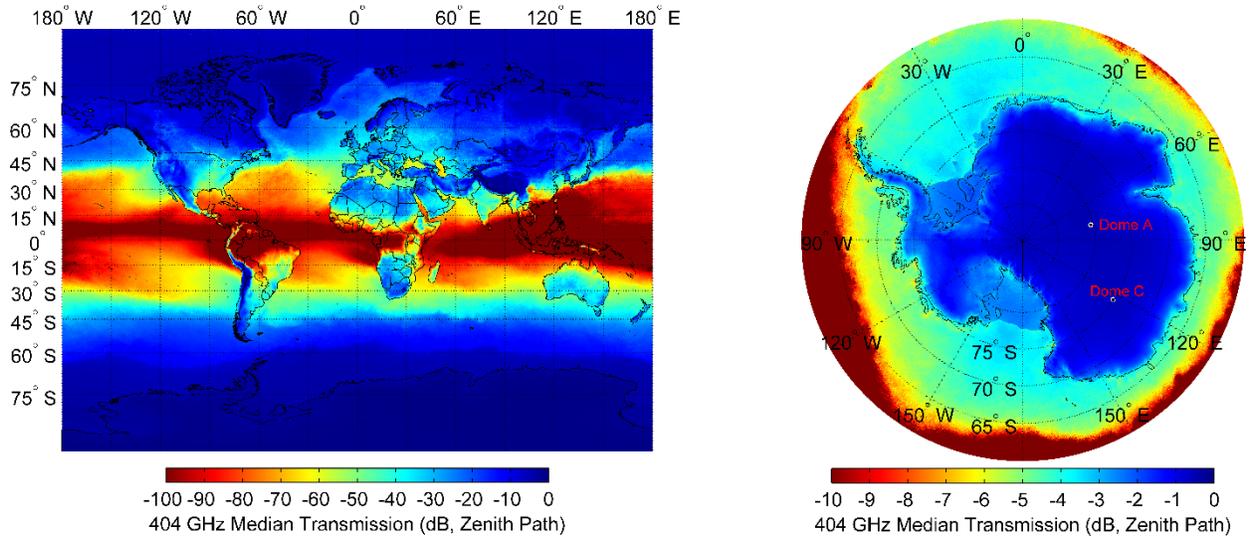

Fig. 11. Median vertical path clear-sky transmission at 404 GHz. The supplemental figures include the world map scaled from -10 to 0 dB.

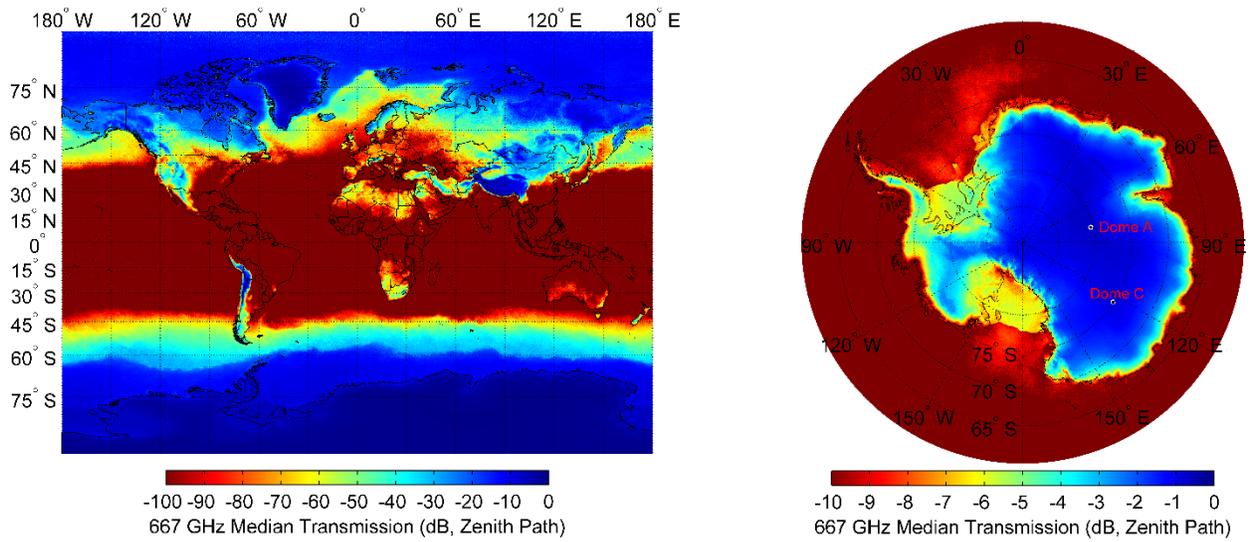

Fig. 12. Median vertical path clear-sky transmission at 667 GHz. The supplemental figures include the world map scaled from -10 to 0 dB.

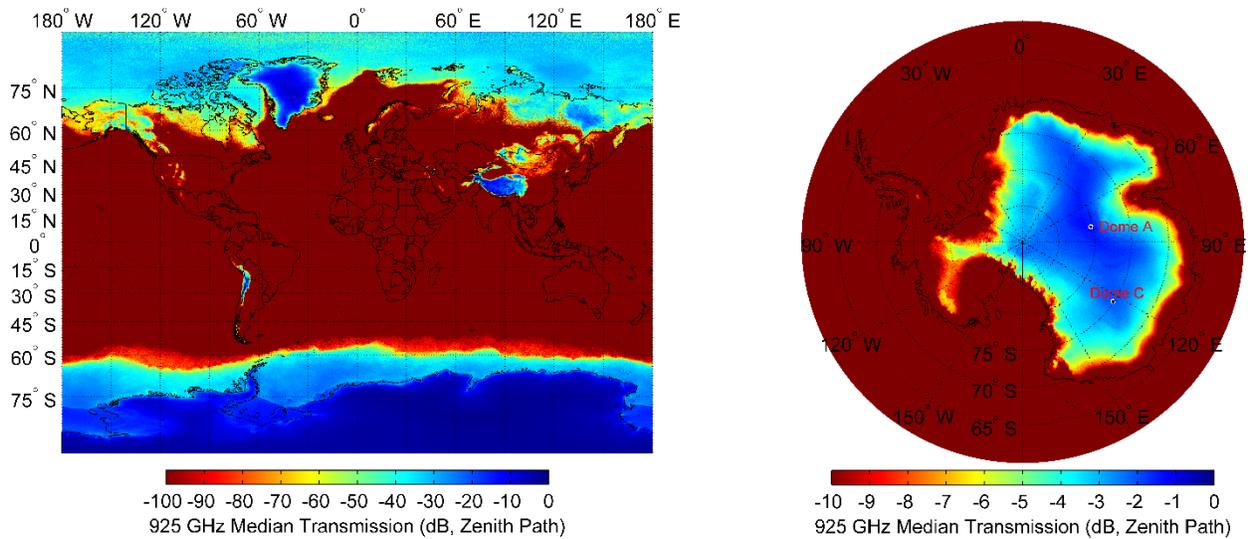

Fig. 13. Median vertical path clear-sky transmission at 925 GHz. The supplemental figures include the world map scaled from -10 to 0 dB.

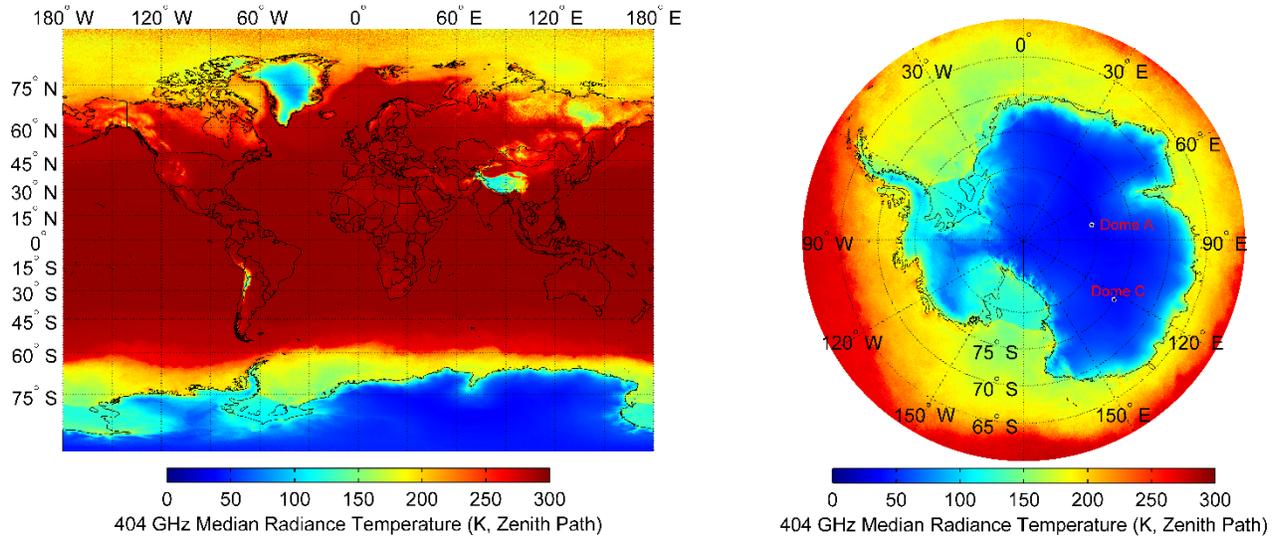

Fig. 14. Median vertical path clear-sky radiance antenna temperature at 404 GHz

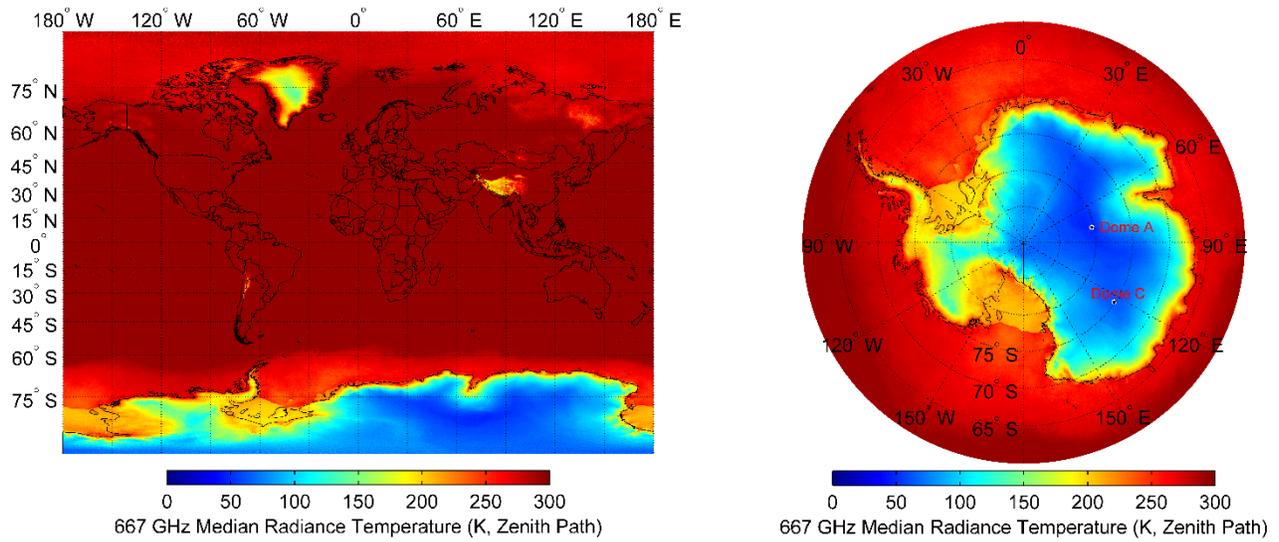

Fig. 15. Median vertical path clear-sky radiance antenna temperature at 667 GHz

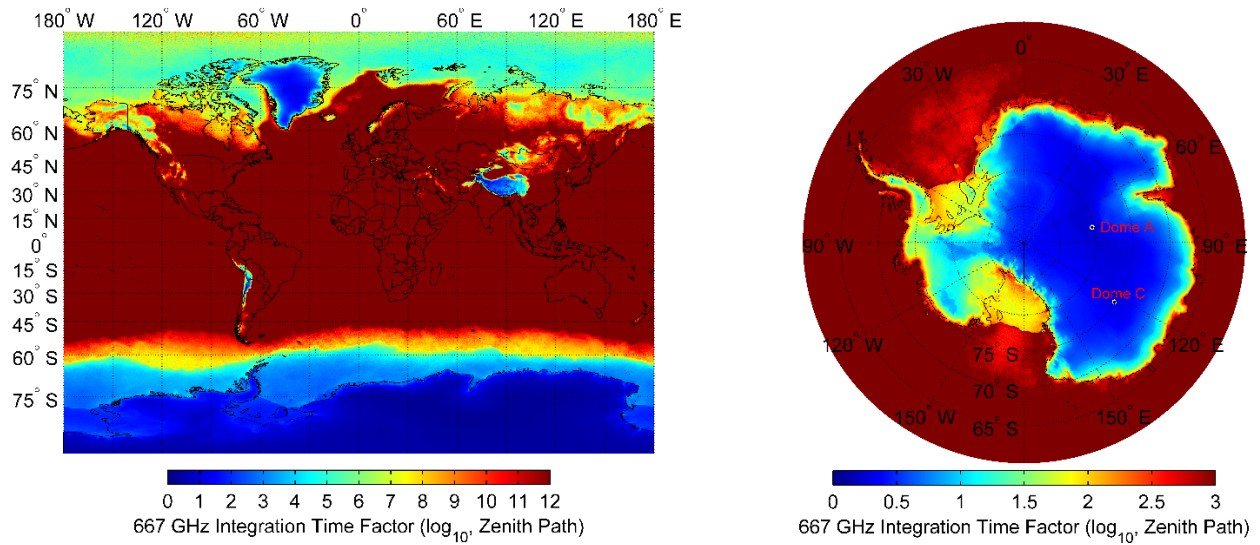

Fig. 16. Vertical path cloud-compensated mean integration time factor at 667 GHz. This factor does not incorporate atmospheric radiance.

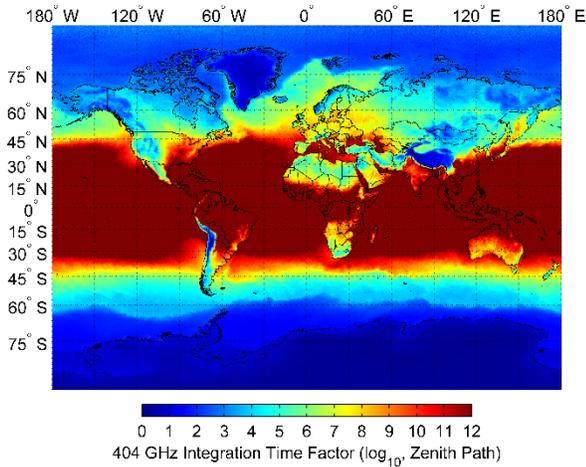

Fig. 17. Vertical path cloud-compensated mean integration time factor at 404 GHz. The lower attenuation in this band results low to moderate integration time factors over larger areas than at 667 GHz (Fig. 13).

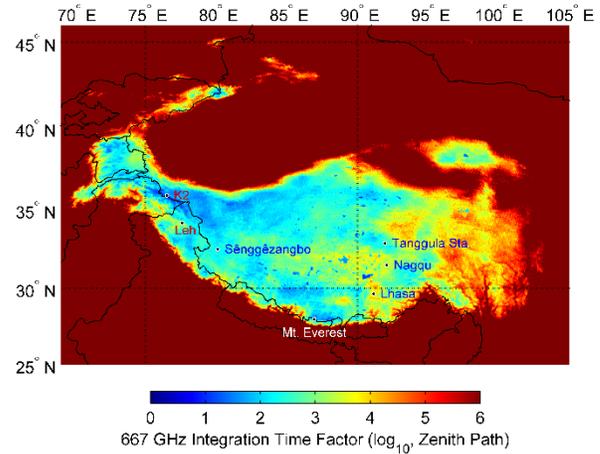

Fig. 18. Vertical path cloud-compensated mean integration time factor at 667 GHz, showing the Tibetan plateau and Himalayas. This factor does not incorporate atmospheric radiance and represents a strong astronomical source or satellite passing overhead.

larger arrays of detectors can be deployed. We assume a factor of 10 to 100 times larger arrays on the ground for a point of discussion. As such, we consider an excess integration time factor of $10^5$ to $10^6$ to be very roughly comparable in terms of long-term productivity to an actual spaceborne telescope. Of course, there are times where this long-term metric does not directly apply, such as the observation of transient phenomena.

The complication of background limits needs to be considered for each case where the noise dominates the signal as well as atmospheric fluctuations, long term stability and systematic errors. The discussion becomes very complex and is highly debated in the community for very low background, weak signal applications.

At 404 GHz (Fig. 17), significantly many more sites over the world are available, including much of the Western United States and the Swiss Alps, with an excess integration time factor of $10^2 - 10^3$. While we ignore radiance for this analysis, for many applications an optimally sited ground based large aperture telescope can be competitive with a smaller space telescope at lower frequencies but again it very much depends on the specifics of the science being done. See our analysis and comparison in [10].

Figure 18 is a detail view of the same data centering on the Tibetan Plateau, where the large area of acceptable sites are visible, along with the proximity to inhabited locations. The city of Lhasa has been noted on the maps. With a population of around 1.4 million, there is far better infrastructure in the region and ease of travel than sites such as Antarctica and the Greenland Ice Sheet and thus is an interesting site worthy of further study.

The Tanggula Railway Station is also noted. Although it is in a currently uninhabited area, it is situated along the Qinghai-Tibet railway and allows for relatively easy site development. Though the Station has a relatively high cloudy sample fraction of over 50 percent, it has a clear-sky PWV superior to that of the ALMA site. In order to compare the relative effects of both, Table II lists the mean cloud-compensated integration time factor for all windows. At 275 GHz the relative effects of PWV and cloud cover are matched. At 404 GHz and above, the Tibetan site excels over the Chilean site. By the fourth THz window at 847 GHz, a telescope sited at Tanggula Station is more than an order of magnitude more productive than an identical one the ALMA site. This once again stresses the fact in the THz bands, low PWV dominates over cloud cover.

TABLE II
COMPARISON OF ALMA SITE WITH TANGGULA STATION

| Location (Altitude) | Integration Time Factor ($\log_{10}$, Cloud-compensated, Zenith Path) | | | | | | | |
|---|---|---|---|---|---|---|---|---|
|  | 275 GHz | 404 GHz | 667 GHz | 847 GHz | 925 GHz | 1015 GHz | 1336 GHz | 1500 GHz |
| ALMA (5050 m) | 0.41 | 1.21 | 2.93 | 3.63 | 7.71 | 15.0 | 16.0 | 15.6 |
| Tanggula (5068 m) | 0.52 | 1.06 | 2.12 | 2.55 | 5.10 | 9.64 | 10.3 | 10.0 |
| Difference | -0.11 | 0.15 | 0.81 | 1.08 | 2.61 | 5.36 | 5.70 | 5.60 |

Comparison of long-term mean integration time factor between the ALMA site and Tanggula Station, Tibet. Tanggula Station is superior to the ALMA site above 275 GHz, where the effects of the lower PWV of the Tibetan site compensate for its higher cloud cover rate. By 847 GHz, the difference is greater than one order of magnitude. (A difference of 1 indicates it takes 10 times longer to integrate and 2 indicates 100 times longer integration versus no atmospheric attenuation.)

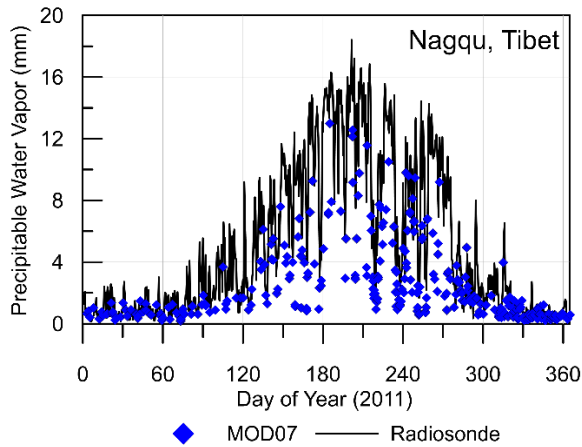

Fig. 19. Time series of Nagqu, Tibet. MOD07 data is compared to twice-daily radiosonde measurements. Seasonality of PWV data is seen from both datasets.

TABLE III
PERFORMANCE OF WELL-KNOWN ASTRONOMICAL SITES FROM MOD07 DATA: QUARTILE STATISTICS

| Location | Clear-Sky PWV by Percentile (mm) | | | All-Weather[1] PWV by Percentile (mm) | | | Median Attenuation (dB, Clear-sky, Zenith Path) | | | | | | | |
|---|---|---|---|---|---|---|---|---|---|---|---|---|---|---|
| | 25th | 50th | 75th | 25th | 50th | 75th | 275 GHz | 404 GHz | 667 GHz | 847 GHz | 925 GHz | 1015 GHz | 1336 GHz | 1500 GHz |
| Dome A, Antarctica | 0.05 | 0.07 | 0.13 | 0.05 | 0.08 | 0.18 | 0 | 0.50 | 0.71 | 0.79 | 1.49 | 2.78 | 2.70 | 2.71 |
| Dome C, Antarctica | 0.07 | 0.10 | 0.19 | 0.08 | 0.11 | 0.24 | 0 | 0.57 | 0.91 | 1.04 | 2.03 | 3.83 | 3.83 | 3.81 |
| South Pole | 0.10 | 0.14 | 0.24 | 0.11 | 0.20 | * | 0.02 | 0.67 | 1.17 | 1.37 | 2.75 | 5.24 | 5.35 | 5.28 |
| Tanggula Station, Tibet | 0.39 | 0.78 | 1.71 | 0.93 | * | * | 0.47 | 2.23 | 5.35 | 6.62 | 14.2 | 27.7 | 29.5 | 28.8 |
| ALMA Site, Chajnantor | 0.86 | 1.37 | 2.20 | 1.0 | 1.86 | * | 0.89 | 3.66 | 9.20 | 11.5 | 24.8 | 48.4 | 51.8 | 50.4 |
| CARMA Site | 3.39 | 5.56 | 9.34 | 3.96 | 7.16 | 20.9 | 3.84 | 13.8 | 36.6 | 45.9 | 99.7 | 196 | 210 | 204 |
| Owens Valley Radio Observatory (OVRO) | 3.49 | 5.36 | 8.87 | 4.36 | 8.57 | * | 3.88 | 14.0 | 37.0 | 46.4 | 101 | 198 | 213 | 207 |

1. All-Weather PWV is estimated by assuming cloudy sample PWV is infinite. Percentiles where the sky is cloudy marked by *.

TABLE IV
PERFORMANCE OF WELL-KNOWN ASTRONOMICAL SITES FROM MOD07 DATA: MEAN STATISTICS

| Location | Cloudy Sample Fraction | Altitude (m) Latitude, Longitude (deg.) | Mean Clear-Sky PWV (mm) | Integration Time Factor ($\log_{10}$, Cloud-compensated, Zenith Path) | | | | | | | |
|---|---|---|---|---|---|---|---|---|---|---|---|
| | | | | 275 GHz | 404 GHz | 667 GHz | 847 GHz | 925 GHz | 1015 GHz | 1336 GHz | 1500 GHz |
| Dome A, Antarctica | 0.138 | 4,091 m -80.36, 77.35 | 0.10 | 0.06 | 0.18 | 0.24 | 0.27 | 0.46 | 0.81 | 0.81 | 0.80 |
| Dome C, Antarctica | 0.112 | 3,223 m -75.1, 123.35 | 0.15 | 0.06 | 0.19 | 0.30 | 0.34 | 0.64 | 1.17 | 1.19 | 1.18 |
| South Pole | 0.267 | 2,835 m -90, 0 | 0.19 | 0.15 | 0.36 | 0.56 | 0.63 | 1.16 | 2.09 | 2.16 | 2.13 |
| Tanggula Station, Tibet | 0.555 | 5,068 m 32.89, 91.92 | 1.31 | 0.52 | 1.06 | 2.12 | 2.55 | 5.10 | 9.64 | 10.3 | 10.0 |
| ALMA Site, Chajnantor | 0.252 | 5,058 m -23.03, -67.75 | 2.10 | 0.41 | 1.21 | 2.93 | 3.63 | 7.71 | 15.0 | 16.0 | 15.6 |
| CARMA Site | 0.198 | 2,196 m 37.28, -118.14 | 7.21 | 1.10 | 3.67 | 9.57 | 12.0 | 26.0 | 50.8 | 54.6 | 53.1 |
| Owens Valley Radio Observatory (OVRO) | 0.321 | 1,222 m 37.23, -118.28 | 7.15 | 1.16 | 3.71 | 9.57 | 12.0 | 25.8 | 50.5 | 54.3 | 52.8 |

We have summarized results for several well-known astronomical observatory sites, as well as Tanggula Station in Tables III and IV. Table III contains metrics based on PWV quartiles, while Table IV is based on mean PWV. As in Table II, the difference in integration time factors between two sites represents the order of magnitude difference in estimated integration time. We again estimated the all-sky percentiles in Table III by assuming the PWV for cloudy samples was infinite. While many clouds have relatively little liquid or gaseous water (as low as 0.2 mm for a ground-level stratus cloud), they can exhibit significant absorption and scattering at THz frequencies, from 15-30 dB/km at 600 GHz [14]. Thus, our estimated all-sky percentiles are perhaps more representative of usable observing time than a true all-sky PWV measurement.

In order to validate the all-sky PWV extrapolation and the measured PWV data in the Tibetian region, we compared MOD07 data to contemperanous radiosonde measurements at Nagqu Town, Tibet (31.48° N, 92.05° E, 4,508 m altitude), obtained from the University of Wyoming weather server[2]. Nagqu is the closest regular radiosonde station to Tanggula Station. Fig. 19 depicts the time series data, showing clear seasonality. Fig. 20 shows the cumulative distribution functions of both the radiosonde and the MOD07 data, scaled for all-sky conditions. Both sources are essentially equal up through the 15th percentile, about 0.67 mm PWV. The differences between the two distributions above this point is likely due to the water vapor content of clouds. We found that in all cases, the MOD07 distribution was equal or wetter than radiosonde distribution, lending strong support to the validity

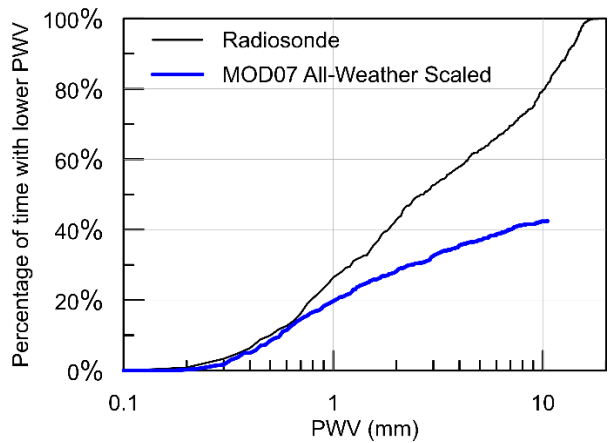

Fig. 20. Distribution of contemporaneous radiosonde and MOD07 data from Nagqu (4,508 m altitude). As in Table III, MOD07 data was scaled to form an all-weather estimate by assuming cloudy samples have infinite PWV. Close correspondence is seen at low PWVs but diverges due to the assumed cloudy-sample PWV. MOD07 data is equal or wetter in all cases, supporting the accuracy of performance estimates of the Tibetian Plateau.

---

[2] http://weather.uwyo.edu/upperair/sounding.html

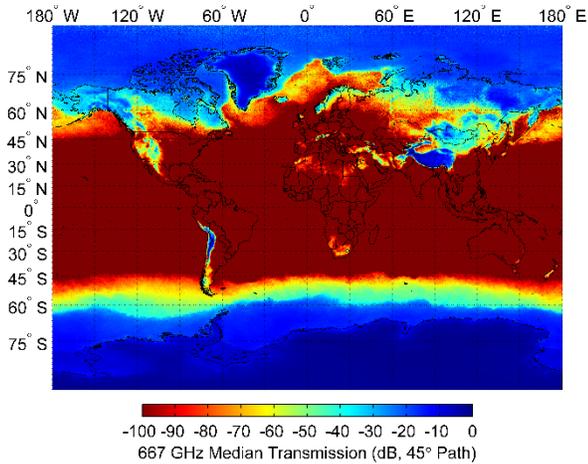

Fig. 21. 45° slant path median clear-sky transmission at 667 GHz. Comparison with Fig. 12 shows the effect of the longer atmospheric propagation.

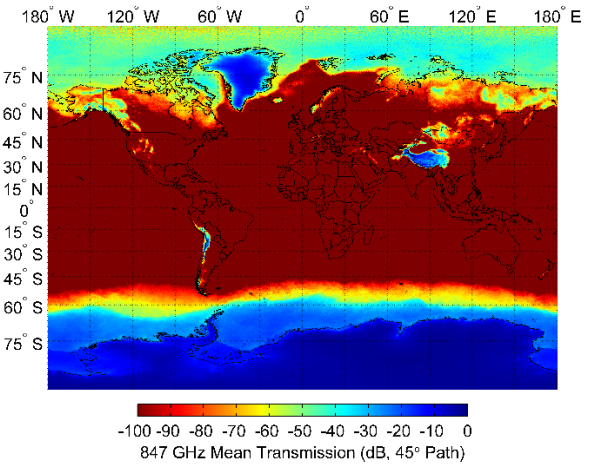

Fig. 22. 45° slant path mean clear-sky transmission at 847 GHz. This scenario approximates a geostationary satellite link.

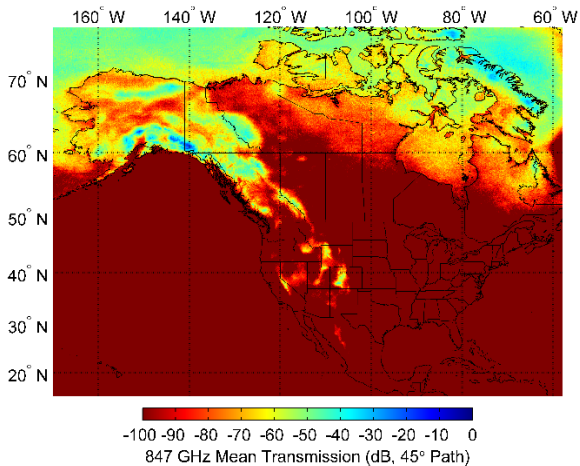

Fig. 23. 45° slant path mean clear-sky transmission at 847 GHz, showing North America. This scenario approximates a geostationary satellite link from mid-latitudes.

to the cloud compensation factors and to the analysis of the Tibetan region.

### B. Communication scenarios

We next turn our attention towards communication scenarios. We used a 45 degree slant path to represent a geostationary satellite link from a mid-latitude site (high equatorial humidity would preclude a link site in that region). Fundamentally the site distribution is similar, with slightly higher attenuation. At the same time, this can represent a lower elevation astronomical observation. Figure 21 shows median clear-sky atmospheric transmission for a 45 degree slant path and can be compared with Fig. 9 to show the effects of the additional path length. In the mid-latitudes, at 667 GHz for example, we see an additional 3-8 dB of loss due to the additional factor of $\sqrt{2}$ in the path length.

We present clear-sky data which is related to operational link data rate. Link availability can be found simply from the cloudy sample fraction. The long-term data rate, that is the number of bits transferred per year, can be estimated by the product of the link rate and the cloudy sample fraction. The integration time factor, taking Fig. 16 as an example, may appear to be very large. It should not, by itself, be interpreted to be a bound or estimate of any data rate. It is incorrect to bound THz data rates by dividing the Shannon limit by the integration time factor, since the former is fundamentally dependent on the signal-to-noise ratio, which can be increased by addition of transmitter power or antenna gain, for example. An estimate of data rates must follow from a full link-budget analysis, which we will undertake in a future work.

Fig. 22 applies the mean transmission model to a 45 degree slant path, accounting for cloud cover, for the 847 GHz band. We see that even at this high frequency, areas with moderate transmission, such as the Scandinavian mountains have attenuations of around 40 dB. We note that for a given antenna, the gain is proportional to $\frac{1}{\lambda^2}$, and thus, on the first order, an increase of frequency from 8.5 GHz to 850 GHz, yields a 40 dB gain. This is undoubtedly a ballpark calculation, but does suggest that, in terms of communication systems, a 40 dB atmospheric attenuation is not an insurmountable challenge to a modern communication system.

Finally, Figure 23 depicts North American sites at 847 GHz and a 45 degree slant path. There are some low attenuation (~20 dB) sites in Alaska, though these are associated with high a high cloudy sample rate (~60%) and rugged terrain. There are a number of medium absorption locations (~40 dB) in the Western United States that suggests the availability of this band for not only commercial space communications, for which the United States is a major consumer of, but for government communication applications where transmission directly to the United States is necessary. We further note that our results strongly suggest that any system using THz frequencies for secure inter-satellite communication use frequencies specifically within a very strong absorption window.

### IV. CONCLUSIONS

Using observed PWV data from the MODIS instrument on board two NASA EOS satellites, we produced a high-resolution map of mean and median PWV and cloud cover data for the 2011 calendar year as well as additional PWV quartiles and temporal distributions. Combined with the MODTRAN atmospheric model, we produced a series of maps depicting transmission and radiance for the 8 atmospheric THz windows. We studied both a best-case zenith-pointing path as well as a 45-degree path, appropriate for a geostationary satellite link. We also combined the cloud cover data and the transmission model to determine the excess integration time factor, representing the additional integration time necessary to reach an identical SNR due solely to atmospheric water vapor for astronomical observations. The continuous, global and high-resolution data provided by the MODIS instrument allowed for unprecedented analysis of worldwide sites.

Although we cannot consider all of the many complex scenarios possible, the data clearly indicated that there are excellent sites that provided acceptable performance for both astronomical observations and satellite communications. The best sites are Antarctica, the Greenland Ice Sheet, the Atacama Desert and the Tibetan Plateau. In the lower frequency THz windows, the peaks of the Western United States, Alaska, and the Scandinavian Mountains offered acceptable transmission. Performance over all of inland Antarctica was quite close at lower frequencies but differentiates significantly at higher frequencies.

Our analysis also show a very wide area of favorable sites on the Tibetan plateau. We feel that this area has been substantially overlooked for THz applications and currently there are no operating THz observatories which exploit this area, nor have major telescopes, similar to ALMA or CCAT have been proposed. We speculate a major reason is due to an unremarkable cloudy sample fraction making it unattractive to optical astronomers. However, when the skies are clear, the very dry atmosphere produces a significant and unique advantage where THz is concerned. The plateau is a harsh environment, yet, unlike the Greenland Ice Sheet, does have numerous settlements, road and rail transportation networks. It further provides a Northern Hemisphere site for celestial observations not possible from the current telescopes on Cerro Chajnantor.

We stress that the siting of any astronomical telescope is often done for specific science goals or a suite of goals and that the difference between sites is a complex function of the specific science requirements. For example, sites with higher PWV (Mauna Kea or White Mountain for example) may be perfectly acceptable for some science goals while extreme sites (Dome A, Ridge A) may be needed for others. It is critical to understand these differences in order to make logical siting decisions. In many cases the issue of atmospheric stability is a complicating factor that is often mitigated by chopping and scanning techniques and not just raw site stability. This is particularly true in some cases where ground, airborne and space missions often have similar science goals and the ultimate limits of the ground based site is often unknown before doing the measurement. This has historically been true of CMB measurements for example.

This is also true in the field of communications where concerns, such as link availability and accessibility of communications infrastructure are critical in siting. Unlike astronomy, designers of satellite links have the luxury of engineering the transmitted signal, and can exploit techniques such as frequency diversity and signal equalization. We support the development of experimental satellite payloads in order to truly characterize THz satellite communication channels.

A significant challenge in the analysis of the data was the sheer volume and need for high performance computing systems. After exploring several other options which did not scale to the necessary performance, we ultimately used MATLAB coupled with a MapReduce parallel programming scheme. While we found this model highly efficient and relatively easy to develop, future, more detailed statistical analysis of the data may require specific algorithms that do not parallelize as well.

While we do not doubt that in the THz regime water vapor poses a fundamental challenge and limits many applications, our work shows that the effects of water vapor attenuation can be significantly mitigated by selection of dry and high-altitude sites, many of which exist over the surface of the Earth. Further research should be undertaken to characterize specific applications, observations, and communication links at these sites which we believe will create new applications and goals for THz systems, sources and detectors. A future paper will expand our analysis of water vapor into the infrared, a region which is also important for both astronomical and optical satellite links.

REFERENCES

[1] J. Marvil, "An astronomical site survey at the Barcroft Facility of the White Mountain Research Station," *New Astron.*, vol. 11, pp. 218-225, Jul. 2005.
[2] S. Radford, "Cerro Chajnantor Water Vapor and Atmospheric Transmission," CCAT Consortium, Ithaca, NY, TM 112, 2012.
[3] P. G. Calisse, et al., "Site testing for submillimetre astronomy at Dome C, Antarctica," *P. Astron. Soc. Aust.*, vol. 21, pp. 256-263, 2004.
[4] E. S. Battstelli, et al., "Intensity and polarization of the atmospheric emission at millimetric wavelengths at Dome Concordia," *Mon. Not. R. Aston. Soc.*, vol. 423, pp. 1293-1299, 2012.
[5] S. De Gregori, et al., "Millimeter and submillimeter atmospheric performance at Dome C, combining radiosoundings and ATM synthetic spectra," *Mon. Not. R. Aston. Soc.*, vol. 425, pp. 222-230, 2012.
[6] Y. Shiao, et al., "Water Vapor in the Atmosphere: An Examination for CARMA Phase Correction," *Proc. SPIE*, vol 6275, no. 62750Y, 2006.
[7] G. Sims, et al., "Where is Ridge A?" *Proc. SPIE*, vol. 8444, no. 84445H-1, 2012.
[8] P. Tremblin, et. al., "Worldwide site comparison for submillimetre astronomy," *Astron. Astrophys.*, vol. 584, no. A65, 2012.
[9] M. R. Swain, et al., "A comparison of possible Antarctic telescope locations," *Proc. SPIE*, vol 6267, no. 62671K, 2006.
[10] S. Denny, et al., "Fundamental Limits of Detection in the Far Infrared," *New Astron.*, accepted, in press, 2013.
[11] V. V. Salomonson, et al., "MODIS: advanced facility instrument for studies of the Earth as a system," *IEEE Trans. Geosci. Remote Sens.*, vol. 27, no. 2, pp. 145-153, 1989.
[12] S. W. Seeman, et al., "MODIS Atmospheric Profile Retrieval Algorithm Theoretical Basis Document, Version 6," Coop. Inst. For Met. Sat. Stud., Madison, WI, 2006.
[13] S. Ackerman, et. al., "Discriminating Clear-sky from Cloud with MODIS Algorithm Theoretical Basis Document (MOD35), Version 5.0," Coop. Inst. For Met. Sat. Stud., Madison, WI, 2006.
[14] H. J. Liebe, et al., "Millimeter-wave attenuation and delay rates due to fog/cloud conditions," *IEEE Trans. Antennas Propag.*, vol. 37, no. 12, 1989.
[15] H. Yang, et al., "Exceptional Terahertz Transparency and Stability above Dome A, Antarctica," *P. Astron. Soc. Pac.*, vol. 122, pp. 490-494, 2010.

[16] P. Tremblin, et. al., "Site testing for submillimetre astronomy at Dome C, Antarctica," *A&A*, vol. 535, no. A112, 2011.
[17] R. A. Chamberlin, et. al., "The wintertime South Pole tropospheric water vapor column: Comparisons of radiosonde and recent terahertz radiometry, use of the saturated column as a proxy measurement, and inference of decadal trends," *J. Geophys. Res.,* vol. 117, D13111, 2012.
[18] J. W. Warner, "Comparative Water Vapor Measurements for Infrared Sites," *P. Astron. Soc. Pac.,* vol. 89, pp 724-727, Oct. 1977.
[19] S. Paine, "The *am* atmospheric model," Smithsonian Astrophys. Obs., Cambridge, MA, Rep. 152, Ver. 7.2, 2012.
[20] J. R. Pardo, et. al., Atmospheric Transmission at Microwaves (ATM): An Improved Model for Millimeter/Submillimeter Applications," *IEEE Trans. Antennas Propag.,* vol. 49, no. 12, 2001.

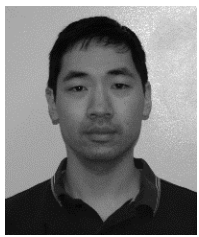

**Jonathan Y. Suen** (SM'03) received an A.A. degree from Bard College at Simon's Rock in 2002, the B.S. degree from the University of California, Santa Barbara in 2004, and the M.S. degree from the University of Michigan, Ann Arbor in 2005, all in Electrical Engineering. He is currently a doctoral candidate at the University of California, Santa Barbara in the Department of Electrical and Computer Engineering.

Since 2008 he has been a graduate student researcher at UCSB. His research centers on the development of terahertz systems and has published research on biomedical imaging, THz sources and components, as well as systems for astrophysical observations. He was the founder of an Internet startup and currently has a patent pending on a novel mm-wave array for long-standoff security imaging.

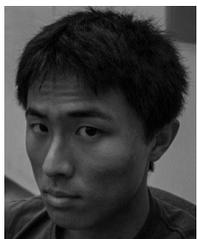

**Michael T. Fang** is currently an undergraduate student pursuing a B.S. degree in Physics and Mathematical Sciences at the University of California, Santa Barbara.

He has been an undergraduate researcher for the UCSB Experimental Cosmology group in the Department of Physics since late 2011 and has been conducting research on millimeter-wave and terahertz astrophysics.

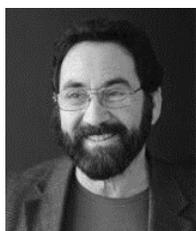

**Philip M. Lubin** received the Ph.D. degree in Physics from the University of California, Bereley, in 1980.

Since 1987 he has been a Professor at the University of California, Santa Barbara. His research centers on millimeter-wave studies of the Cosmic Microwave Background (CMB). He is involved with the COBE and Planck cosmology satellites as well as leading the development of ground- and balloon-borne instruments. His research also includes work on THz and infrared detectors and systems.

Prof. Lubin was a co-recipient of the Gruber Prize in Cosmology in 2007 for his work on the Cosmic Background Explorer (COBE) satellite mission on ground-breaking studies of the CMB.